\documentclass[manuscript,trackchanges]{aastex63}

\usepackage{lineno}

\usepackage[]{natbib}

\usepackage{epsfig}

\usepackage{graphicx}

\usepackage{epstopdf}

\usepackage{amssymb}

\usepackage{lineno}

\usepackage{color,soul}

\usepackage{graphics}

\usepackage{graphics}

\received{December 11, 2019}
\revised{January 23, 2020}
\accepted{February 1, 2020}

\submitjournal{AJ}

\shorttitle{Sanchez et al. 2020}
\shortauthors{Sanchez et al. 2020}

\begin{document}



\title{A new method for deriving composition of S-type asteroids from noisy and incomplete near-infrared spectra}

\correspondingauthor{Juan A. Sanchez}
\email{jsanchez@psi.edu}

\author{Juan A. Sanchez}
\affiliation{Planetary Science Institute, 1700 East Fort Lowell Road, Tucson, AZ 85719, USA}

\author{Cristina Thomas}
\affiliation{Department of Astronomy and Planetary Science, Northern Arizona University, PO Box 6010, Flagstaff, AZ 86011, USA}

\author{Vishnu Reddy}
\affiliation{Lunar and Planetary Laboratory, University of Arizona, 1629 E University Blvd, Tucson, AZ 85721-0092}

\author{Noah Frere}
\affiliation{Department of Physics \& Astronomy, University of Tennessee, 1408 Circle Drive, Knoxville, TN 37996, USA}

\author{Sean S. Lindsay}
\affiliation{Department of Physics \& Astronomy, University of Tennessee, 1408 Circle Drive, Knoxville, TN 37996, USA}

\author{Adriana Mitchell}
\affiliation{Department of Aeronautics and Astronautics, Massachusetts Institute of Technology, 125 Massachusetts Ave, Cambridge, MA 02139, USA}

\begin{abstract}

The surface composition of S-type asteroids can be determined using band parameters extracted from their near-infrared (NIR) spectra (0.7-2.50 $\mu$m) along with spectral calibrations derived from laboratory samples. In the 
past, these 
empirical equations have been obtained by combining NIR spectra of meteorite samples with information about their composition and mineral abundance. For these equations to give accurate results, the characteristics 
of the laboratory spectra they are derived from should be similar to those of asteroid spectral data (i.e., similar signal-to-noise ratio (S/N) and wavelength range). Here we present new spectral calibrations that can 
be used to determine the mineral composition of ordinary chondrite-like S-type asteroids. Contrary to previous work, the S/N of the ordinary chondrite spectra used in this study has been decreased to recreate the 
S/N typically observed among asteroid spectra, allowing us to obtain more realistic results. In addition, the new equations have been derived for five wavelength ranges encompassed between 0.7 and 2.50 $\mu$m, making it possible 
to determine the composition of asteroids with incomplete data. The new spectral calibrations were tested using band parameters measured from the NIR spectrum of asteroid (25143) Itokawa, and comparing the results 
with laboratory measurements of the returned samples. We found that the spectrally derived olivine and pyroxene chemistry, which are given by the molar contents of fayalite (Fa) and ferrosilite (Fs), are in excellent 
agreement with the mean values measured from the samples (Fa$_{28.6\pm1.1}$ and Fs$_{23.1\pm2.2}$), with a maximum difference of 0.6 mol\% for Fa and 1.4 mol\% for Fs.

\end{abstract}

\keywords{minor planets, asteroids: general}

\section{Introduction}

Understanding the compositions of asteroids is crucial to answering important questions regarding solar system formation and evolution. Due to our ability to remotely determine the compositions of many asteroids, studies can 
span from detailed investigations of individual objects to questions regarding the compositional distribution of asteroids throughout the solar system. To determine the compositions of asteroids, researchers seek diagnostic spectral 
band parameters, which enable them to investigate their mineralogies remotely using telescopes. 

For objects belonging to the S, Q, and V classes, which show prominent 1 and 2 $\mu$m absorption features, spectral band parameters such as band centers can be used to determine a more precise composition than 
taxonomy alone provides \citep[e.g.,][]{1972Moon....4...93B, 1974JGR....79.4829A, 1993Icar..106..573G, 2012Icar..220...36S, 2013Icar..222..273D, 2014Icar..228..217T}. These diagnostic absorption features are indicative of 
crystalline olivine ((Mg, Fe)$_2$SiO$_4$) and/or pyroxene ((Mg, Fe)$_2$Si$_2$O$_6$). Olivine characteristically shows a broad absorption feature centered near 1.04-1.1 $\mu$m that is comprised of three overlapping bands, 
while pyroxenes show two broad absorption features centered at 0.9-1 $\mu$m and 1.9-2 $\mu$m. For olivine-pyroxene mixtures, the wavelength of the 1 $\mu$m feature (Band I, BI) is a function of the relative abundances and 
compositions of olivine and pyroxene, while the wavelength of the 2 $\mu$m feature (Band II, BII) is a function of the pyroxene composition \citep[e.g.,][]{1986JGR....9111641C}. The ratio of the area of Band II to Band I, or band 
area ratio (BAR), is also a measure of the relative abundances of olivine and pyroxene.

The relationships between band parameters and mineralogical composition can be rigorously calculated for various types of meteorites, including ordinary chondrites \citep{2002aste.conf..183G, 2010Icar..208..789D}, 
olivine-dominated meteorites \citep{2014Icar..228..288S}, primitive achondrites \citep{2019M&PS...54..157L}, and basaltic achondrites \citep{2009M&PS...44.1331B, 2018JGRE..123.1791B}. \citet{2010Icar..208..789D, 2010M&PS...45..123D, 2010M&PS...45..135D} used a combination of band parameters from laboratory spectra of meteorites, measured mineral abundances from x-ray diffraction (XRD), and mafic silicate compositions determined from electron 
microprobe analysis to formulate the mathematical relationships between band parameters and composition for ordinary chondrite meteorites. These quantitative relationships can be applied to asteroids whose spectra are similar to 
ordinary chondrites \citep[e.g., the S(IV) region defined by][]{1993Icar..106..573G}. 

The current method of determining the composition of ordinary chondrite-like asteroids depends on these laboratory studies that correlate meteorite mineralogies with spectral band parameters from high signal-to-noise ratio (S/N) 
laboratory spectra. The accuracy and precision of any given calculation of asteroid composition depends greatly on the S/N of the observed asteroid spectrum and the researcher’s ability to duplicate the methodology with which 
band parameters are determined. Many works \citep[e.g.,][]{2012Icar..220...36S, 2014Icar..228..217T} also incorporate corrections for temperature and phase angle of the asteroid during observation to adjust the band parameters 
such that they match the conditions of the laboratory measurements. 

There are several factors that hinder our ability to accurately apply these derived mineralogical relationships to near-infrared (NIR) spectra of asteroids. The use of a dichroic filter in an NIR spectrograph can exclude the local 
maximum at $\sim$0.74 $\mu$m that is used to define the continuum for Band I. In the past, these observations have been complemented by visible wavelength spectra ($\sim$0.4-0.9 $\mu$m) from legacy surveys such as 
SMASS \citep[e.g.,][]{1995Icar..115....1X,2002Icar..158..106B}. However, the majority of new asteroid spectral observations in the last two decades are only in NIR wavelengths. Noise from incomplete correction of telluric bands at 
1.8-1.9 $\mu$m can impede our ability to accurately measure the Band II center. At the long wavelength end of the spectrum ($\sim$2.5 $\mu$m), a drop in the quantum efficiency of the detector can lead to a decrease in S/N. For 
Band II, the long wavelength end is often nominally defined as the end of the spectrum. The increased noise at $\sim$2.5 $\mu$m makes it difficult to constrain the linear continuum that the band parameter calculations depend 
upon. Researchers strive to find a way to remain consistent with the methodology used to calculate band parameters for the high S/N meteorite spectra that the composition derivations are defined by, but the increased noise 
makes it hard to determine the true end of the absorption band. Adding uncertainty to the positioning of the continuum can have large effects on the band parameters themselves. 

In this work, we present a new set of spectral calibrations for determining the mineral composition and abundance of ordinary chondrite-like S-type asteroids using the measured compositions and spectra of meteorites from 
\citet{2010Icar..208..789D, 2010M&PS...45..123D, 2010M&PS...45..135D}. The meteorite spectra were altered to simulate the wavelength coverage and S/N of ground-based asteroid observations. We decreased the S/N of the 
spectra using artificially generated errors designed to imitate the error profile of actual observations. We also introduced data cutoffs at 0.8 $\mu$m to simulate the short wavelength edge of NIR spectral observations with a dichroic 
filter and at 2.4, 2.45, and 2.5 $\mu$m to reproduce removing the noisier data at the long wavelength end. Using the altered spectral data, we define a process to determine mineralogy for ordinary chondrite-like S-type asteroids 
under realistic ground-based observing conditions.

\section{The Sample}

For this study we used the same sample described in \citet{2010Icar..208..789D, 2010M&PS...45..123D, 2010M&PS...45..135D}, which consists of a total of 48 equilibrated ordinary chondrites (petrologic types 4-6) belonging 
to the three different subtypes, LL, L, and H.  Modal abundances of 13 LL, 17 L, and 18 H chondrites were measured by \citet{2010M&PS...45..123D} using XRD. Olivine and low-Ca pyroxene compositions, which are given by the 
molar contents of fayalite (Fa) and ferrosilite (Fs), were only measured for 38 of the samples using a Cameca SX-50 electron microprobe \citep{2010M&PS...45..135D}. Samples were ground into fine powders and 
sieved to a grain size of $<$ 150 $\mu$m \citep{2010Icar..208..789D}. Visible and near-infrared (VNIR) spectra (0.32-2.55 $\mu$m) of the 48 samples were acquired at the Reflectance Experiment Laboratory (RELAB) using a 
bidirectional spectrometer. An emission angle of 0$^\mathrm{o}$ and an incident angle of 30$^\mathrm{o}$ were used for all measurements \citep{2003AMR....16..185B, 2010Icar..208..789D}.

\section{Analysis}

\begin{figure*}[!ht]
\begin{center}
\includegraphics[height=10cm]{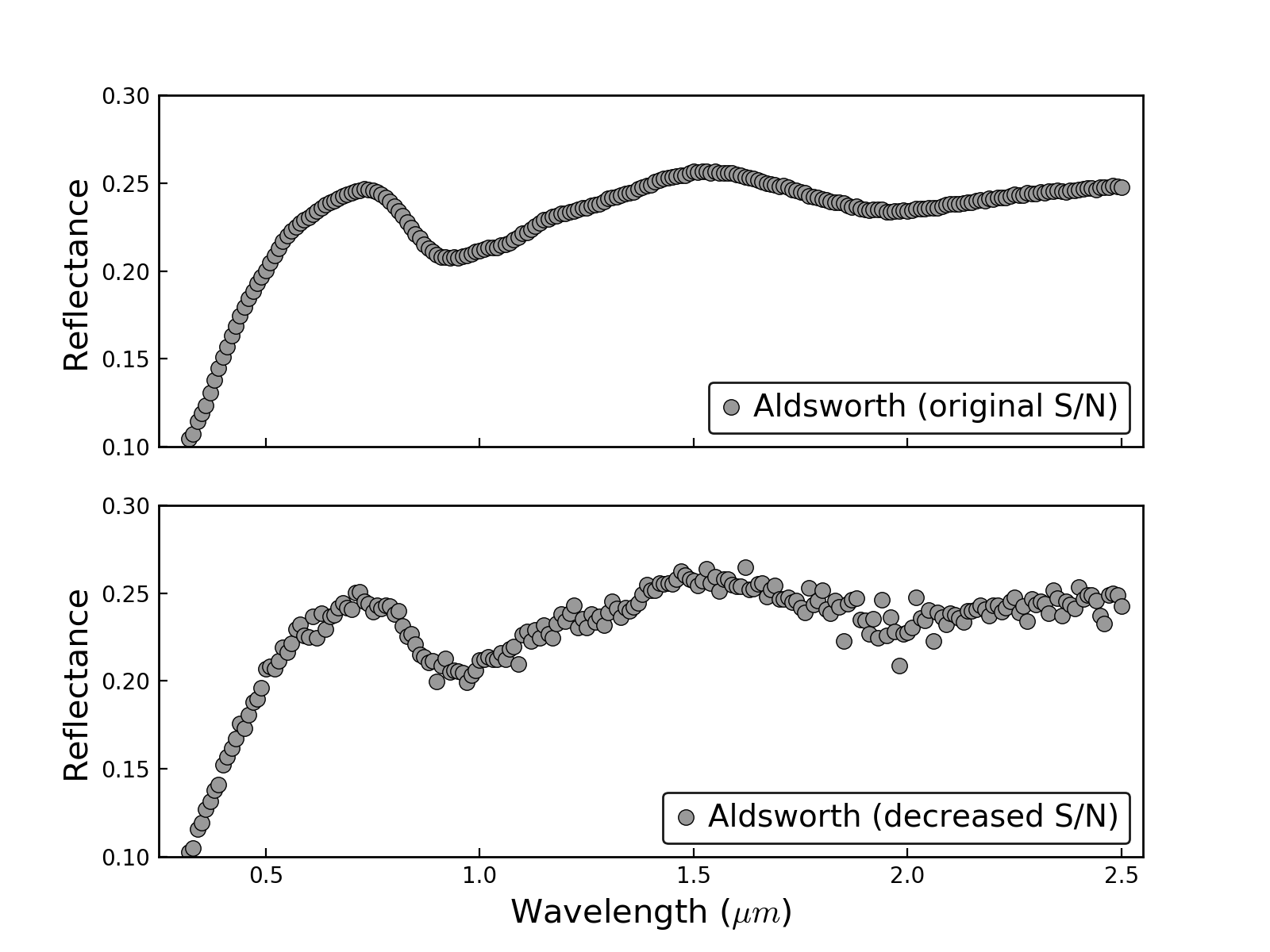}

\caption{\label{f:Figure1} {\small Top panel: original spectrum of the LL5 chondrite Aldsworth with an S/N of $\sim$ 600. Bottom panel: Aldsworth after decreasing the S/N of the entire spectrum to $\sim$ 50, and to $\sim$ 25 for 
wavelengths 1.8-2.1 $\mu$m to simulate the effect of the telluric bands.}}

\end{center}
\end{figure*}

The first step in our analysis is to decrease the S/N of the ordinary chondrite spectra in order to reproduce the value measured for asteroid spectra. For this, we have chosen an S/N of $\sim$ 50, which is typically in the lower limit 
of NIR spectra obtained with the SpeX instrument on the NASA Infrared Telescope Facility (IRTF).  The new degraded spectra were created with a Python code that makes use of the {\it{numpy.random.normal}} function 
\citep[e.g.,][]{2011CSE....13b..22V}. With this function it is possible to generate random samples from a normal (Gaussian) distribution, where the 1$\sigma$ used to create the random samples is calculated from the S/N that we 
choose, in this case $\sim$ 50 for the whole spectrum and about half of that value for the region between 1.8 and 2.1 $\mu$m to simulate the effect of the telluric bands on the spectra. The S/N of the new spectra was then verified 
using the {\it{estimateSNR}} function of {\it{PyAstronomy}}\footnote{https://github.com/sczesla/PyAstronomy} \citep{2019ascl.soft06010C}. Figure \ref{f:Figure1} shows an example for the LL5 ordinary chondrite Aldsworth. The original 
spectrum (top panel) has an S/N of $\sim$ 600, while the S/N of the new spectrum has been decreased about 10 times (bottom panel).

\begin{figure*}[!ht]
\begin{center}
\includegraphics[height=10cm]{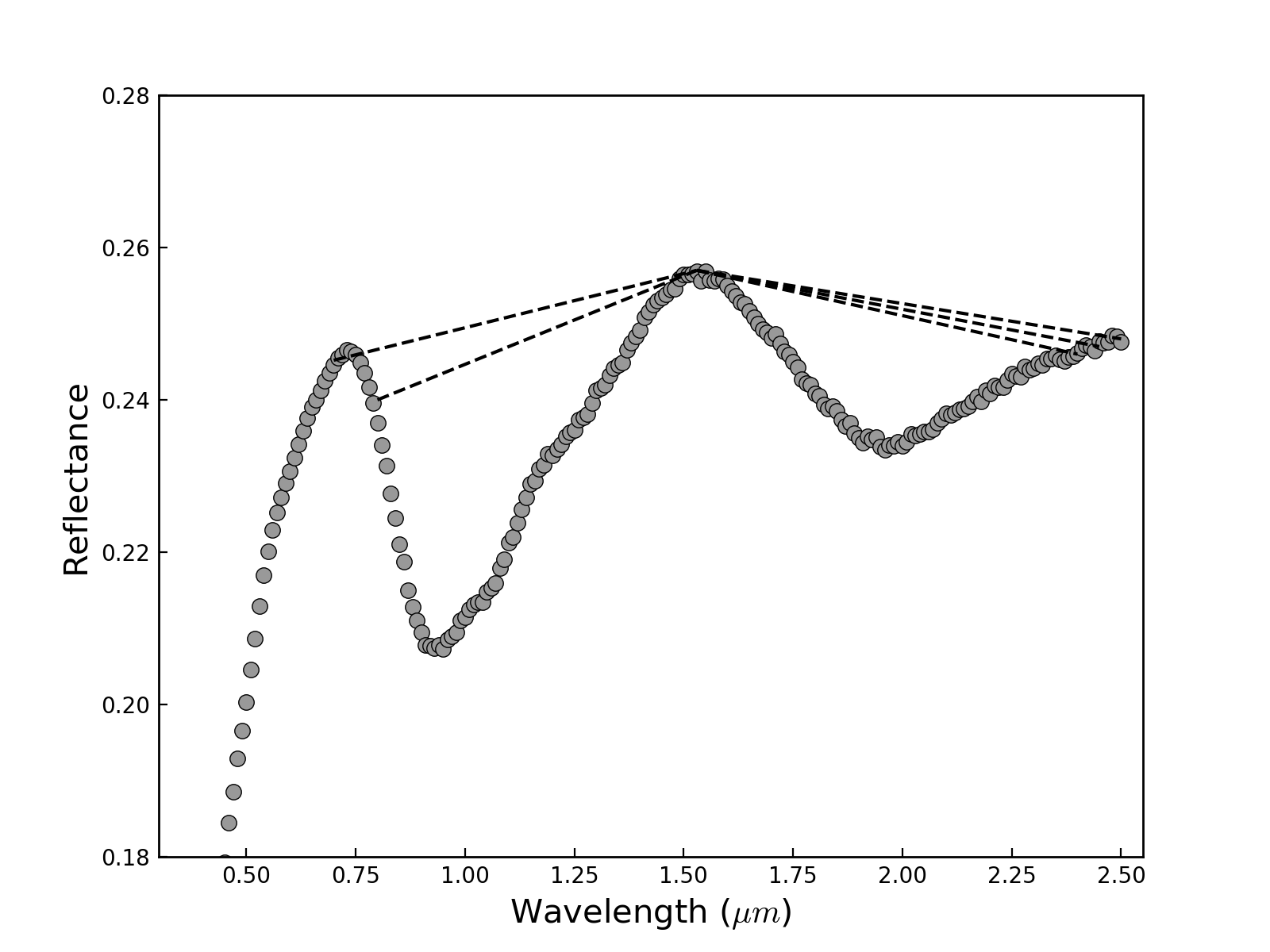}

\caption{\label{f:Figure2} {\small Laboratory spectrum showing the five different wavelength ranges used in this study: 0.7-2.40 $\mu$m, 0.7-2.45 $\mu$m, 0.7-2.50 $\mu$m, 0.8-2.40 $\mu$m, and 0.8-2.45 $\mu$m.}}

\end{center}
\end{figure*}

Spectral band parameters were measured using a Python code similar to the one used in \cite{2015ApJ...808...93S, 2017AJ....153...29S}. Five different wavelength ranges were used in this study: 0.7-2.40 $\mu$m, 0.7-2.45 $
\mu$m, 0.7-2.50 $\mu$m, 0.8-2.40 $\mu$m, and 0.8-2.45 $\mu$m (Figure \ref{f:Figure2}). The short wavelength edge of 0.7 $\mu$m was chosen to include the local maximum at $\sim$0.74 $\mu$m, that allow us to measure the 
continuum for Band I. The cutoff at 0.8 $\mu$m was included to simulate the short wavelength edge of NIR spectra obtained with a 0.8 $\mu$m dichroic filter, such as does obtained with the SpeX instrument on NASA IRTF prior to 
2017. The cutoffs at 2.40 and 2.45 $\mu$m were chosen to account for the decreased response of the detector for wavelengths beyond 2.4 $\mu$m. The wavelength range of 0.7-2.50 $\mu$m was included mostly to compare our results with those of \cite{2010Icar..208..789D}. Although  \cite{2010Icar..208..789D} used spectra in the range of 0.32-2.55 $\mu$m, the effective wavelength range used to extract the band parameters was $\sim$0.7-2.50 $\mu$m. The procedure used to measure the band parameters for the five wavelength ranges was different. For the three wavelength ranges 
encompassed between 0.7 and 2.50 $\mu$m, reflectance maxima were determined by fitting third-order polynomials at $\sim$0.74 and $\sim$1.4 $\mu$m, and a straight line from 2.20 to 2.40, 2.45, and 2.50 $\mu$m. The linear 
continuum was then fitted from the reflectance maxima at 0.74 to the reflectance maxima at 1.4 $\mu$m, and from 1.4 $\mu$m to the three different wavelength ends. After dividing out the linear continuum, band centers were 
calculated by fitting a polynomial over the bottom third of Band I and bottom half of Band II. For the Band I center, 50 measurements were obtained by sampling slightly different wavelength ranges of data points. This was done for 
two consecutive polynomial orders ranging from second to fourth order. This is because we noticed that for some spectra higher polynomial orders yield a better fit. The final value is given by the average of the 100 measurements 
obtained for the two different polynomial orders, e.g., the average of a second and a third or, if higher polynomials were required to obtain a better fit, the average of a third and a fourth. The Band II center was calculated by taking 
the average value of a second and a third polynomial order in all cases, since the shape of the Band II is not as complex as the Band I, and in general shows more scattering due to an incomplete correction of the telluric bands.  The 
BAR is calculated as the ratio of the area of Band II to that of Band I. Band areas are defined as the area between the linear continuum and the data curve, and are calculated using trapezoidal numerical 
integration. Like the Band centers, 100 measurements were done, but in this case, by slightly varying the position where the linear continuum was fitted, the average of these measurements was taken to obtain the final value.  
Errors associated with these parameters are given by the standard deviation of the mean. For the 0.8-2.40 and 0.8-2.45 $\mu$m wavelength ranges, the procedure to measure the band parameters was very similar, the main 
differences being that the reflectance maximum at 0.8 $\mu$m was determined by fitting a straight line from 0.8 to 0.86 $\mu$m, and the Band I center was calculated by fitting a polynomial over the bottom half of the Band I, since 
the area of this band was now smaller than in the previous case.

\section{Results}

\subsection{Olivine-pyroxene abundance ratio}

\begin{figure*}[!ht]
\begin{center}
\includegraphics[height=8.5cm]{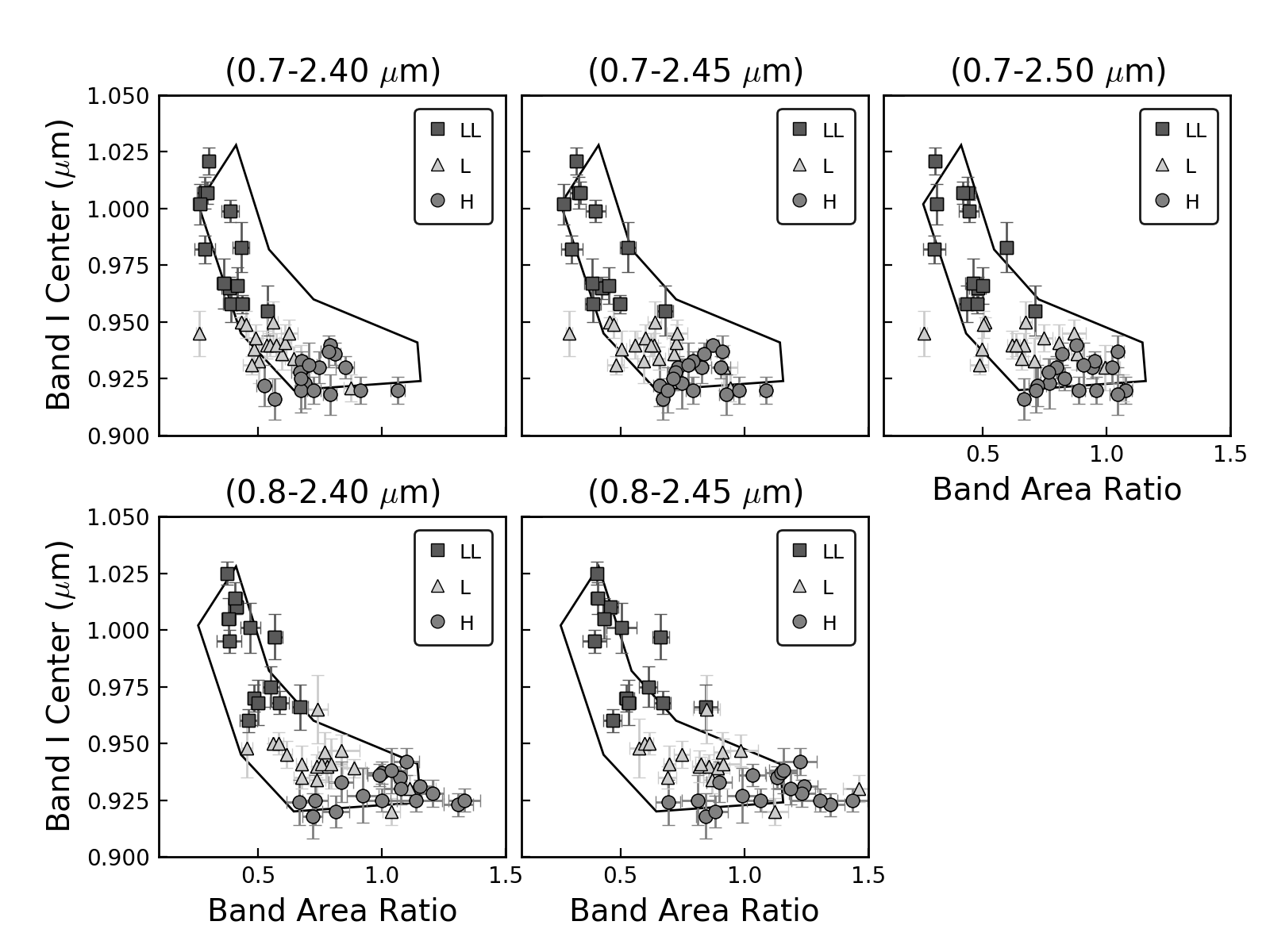}

\caption{\label{f:Figure3} {\small Band I center vs. BAR measured for the three different subtypes of ordinary chondrites for five different wavelength ranges. The polygonal region corresponds to the S(IV) subgroup of 
\cite{1993Icar..106..573G}.}}

\end{center}
\end{figure*}

Figure \ref{f:Figure3} shows the Band I center vs. BAR measured for the LL, L, and H ordinary chondrites for the five different wavelength ranges. The polygonal region corresponding to the S(IV) subgroup 
of \cite{1993Icar..106..573G} is also indicated. Overall, we noticed that our Band I centers are shifted to shorter wavelengths compared to the values measured by \cite{2010Icar..208..789D}. For example, for the 0.7-2.5 $\mu$m 
wavelength range \citep[the same used by][]{2010Icar..208..789D}, there is a difference of $\sim$ 0.02 $\mu$m in the Band I center for the LL and L, and a difference of $\sim$ 0.01 $\mu$m for the H chondrites. This difference is 
mostly the result of the polynomials used to determine the band centers. \cite{2010Icar..208..789D} only used a second-order polynomial, whereas we used second- to fourth-order polynomials. Our results are consistent with the 
findings of \cite{2020Icar..336..113426M}. 

\begin{figure*}[!ht]
\begin{center}
\includegraphics[height=8.5cm]{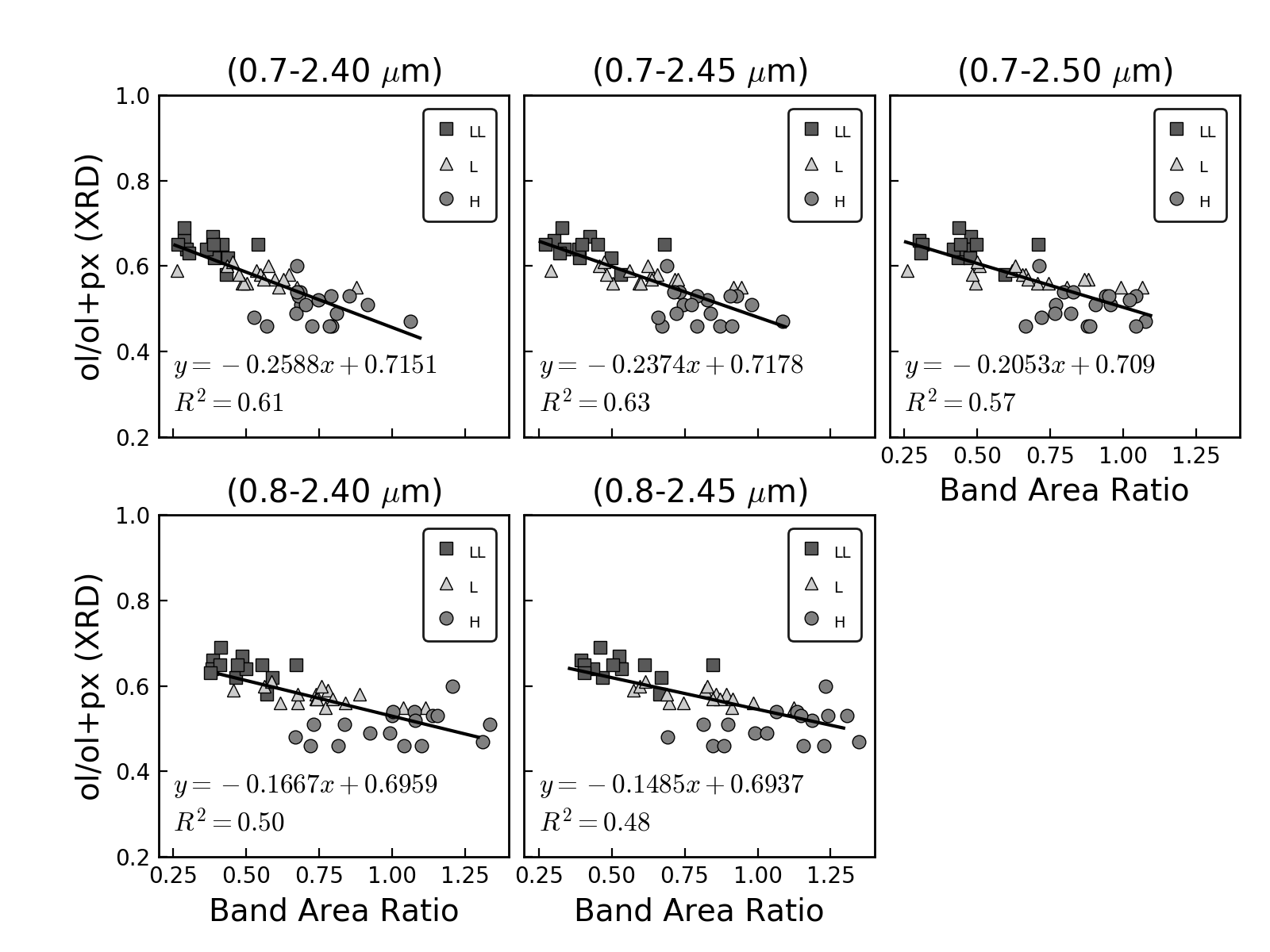}

\caption{\label{f:Figure4} {\small XRD-measured ol/(ol+px) ratio vs. BAR for five different wavelength ranges. A least-square fit of the data and coefficient of determination R$^{2}$ are shown.}}

\end{center}
\end{figure*}

In the case of the BAR, for the three wavelength ranges encompassed between 0.7 and 2.50 $\mu$m, values will shift to the left as the furthest data point moves from 2.50 to 2.40 $\mu$m. The 
same happens for the two wavelength ranges encompassed between 0.8 and 2.45 $\mu$m, however because the 1 $\mu$m band was truncated at 0.8 $\mu$m, BAR values are larger compared to those measured in the range of 
0.7-2.50 $\mu$m. As a result, some H chondrites are shifted to the right and now fall in the S(VI) subgroup (BAR $\sim$ 1.2-1.5) of \cite{1993Icar..106..573G}.

Following the same procedure used by \cite{2010Icar..208..789D} we derived new equations for determining the olivine-pyroxene abundance ratio (ol/(ol+px)) using the BAR values and the XRD-measured modal abundances. Figure 
\ref{f:Figure4} shows the XRD-measured ol/(ol+px) ratio vs. BAR and the linear fits for the five different wavelength ranges. The new spectral calibrations are presented in Table 1. For the five equations, the root mean square error 
(rms) of the spectrally derived ol/(ol+px) ratios is 0.04. The rms is slightly higher than the uncertainty of 0.03 obtained by \cite{2010Icar..208..789D} for the original equation. We found that the coefficient of determination (R$^{2}$) 
varies from 0.48 (worse case corresponding to the 0.8-2.45 $\mu$m range) to 0.63 (best case for the 0.7-2.45 $\mu$m range).

\subsection{Iron abundance in silicate minerals}

\begin{figure*}[!ht]
\begin{center}
\includegraphics[height=10cm]{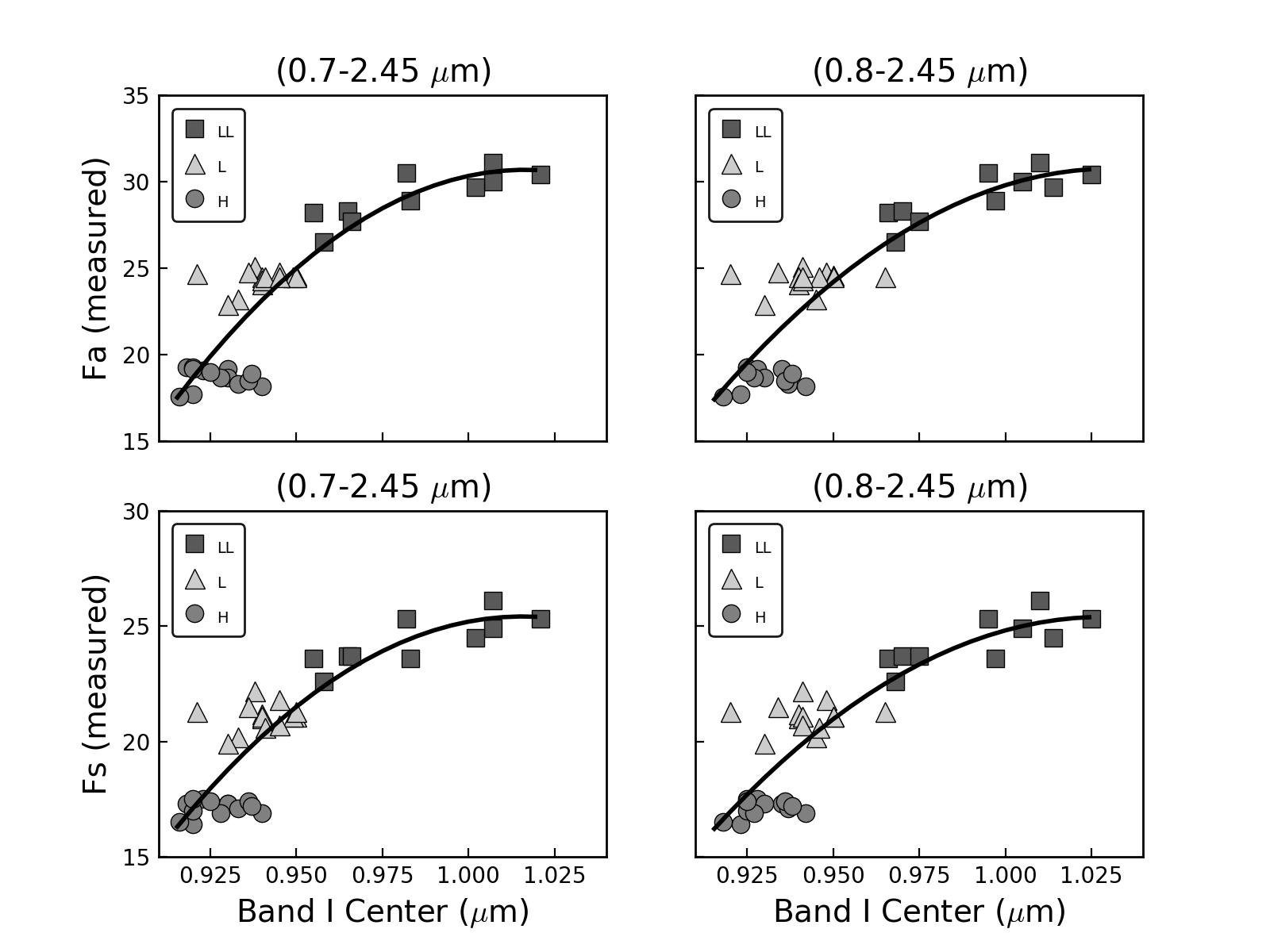}

\caption{\label{f:Figure5} {\small Top panel: measured mol\% of fayalite (Fa) vs. Band I center for wavelength ranges 0.7-2.45 and 0.8-2.45 $\mu$m. Bottom panel: measured mol\% of ferrosilite (Fs) vs. Band I center for wavelength 
ranges 0.7-2.45 and 0.8-2.45 $\mu$m. The Band I center is not affected by changes in the long wavelength end, thus the same equation can be used for the three wavelength ranges encompassed between 0.7 and 2.50 $\mu$m, 
and the two wavelength ranges encompassed between 0.8 and 2.45 $\mu$m.}}

\end{center}
\end{figure*}

\cite{2010Icar..208..789D} found a strong correlation between the Band I center and the iron abundance in olivine (Fa) and pyroxene (Fs). Similarly to \cite{2010Icar..208..789D}, we found that this correlation can be described by a 
second-order polynomial fit (Figure \ref{f:Figure5}). 

The Band I center is not affected by changes in the long wavelength end, thus the same equation  can be used for the three wavelength ranges encompassed between 0.7 and 2.50 $\mu$m. An example for the 0.7-2.45 $\mu$m 
wavelength range is shown in Figure \ref{f:Figure5} (left panels). We observed some variations in the Band I center when the short wavelength edge was changed from 0.7 to 0.8 $\mu$m. This is due to a difference in the 
spectral slope of the continuum that occurs when the reflectance maximum at the short wavelength edge is changed. Therefore, equations for the two wavelength ranges encompassed between 0.8 and 2.45 $\mu$m were also 
derived (Figure \ref{f:Figure5}, right panels). The new equations to determine the mol\% of Fa and Fs are included in Table 1. The rms errors for Fa and Fs were found to be 2.0 and 1.4, respectively. 

\begin{figure*}[!ht]
\begin{center}
\includegraphics[height=10cm]{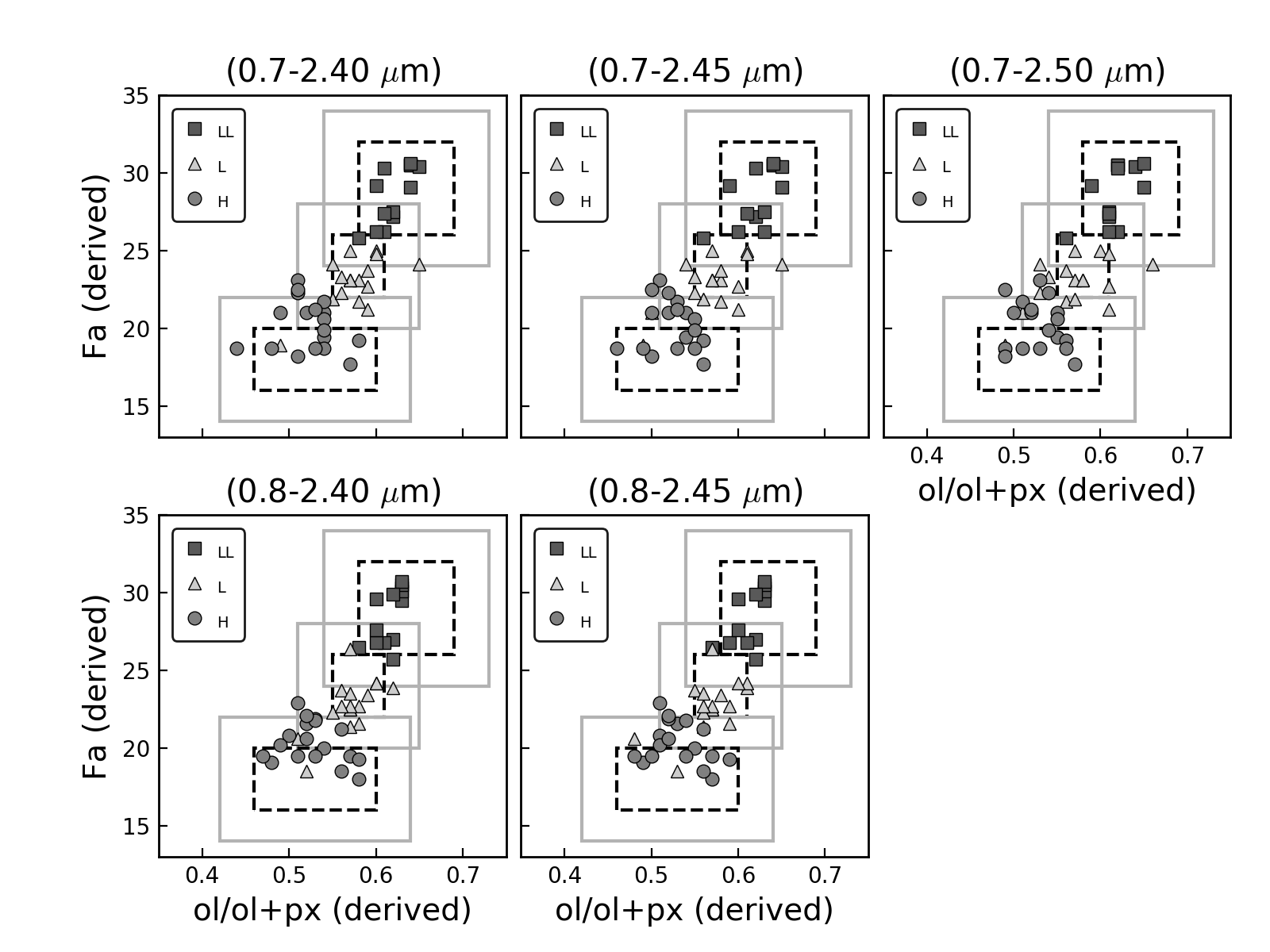}

\caption{\label{f:Figure6} {\small Molar content of fayalite (Fa) vs. ol/(ol+px) ratio derived for LL, L, and H ordinary chondrites for five different wavelength ranges. Black dashed boxes represent the range of measured values for each ordinary chondrite subgroup. Gray solid boxes correspond to the uncertainties associated with the spectrally derived values. Figure adapted from \cite{2010Icar..208..789D}.}}

\end{center}
\end{figure*}

\begin{figure*}[!ht]
\begin{center}
\includegraphics[height=10cm]{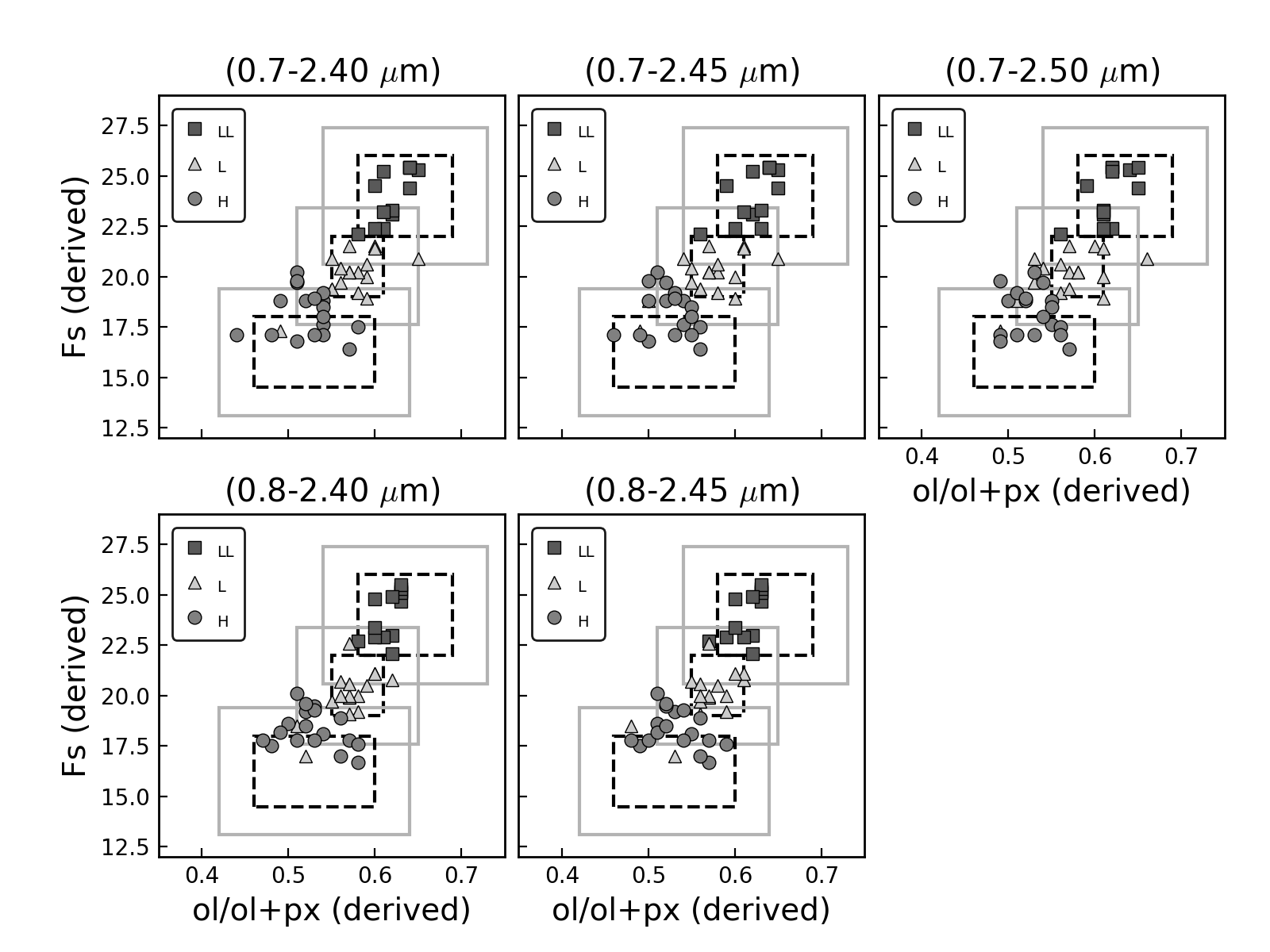}

\caption{\label{f:Figure7} {\small Molar content of ferrosilite (Fs) vs. ol/(ol+px) ratio derived for LL, L, and H ordinary chondrites for five different wavelength ranges. Black dashed boxes represent the range of measured values for each ordinary chondrite subgroup. Gray solid boxes correspond to the uncertainties associated with the spectrally derived values. Figure adapted from \cite{2010Icar..208..789D}.}}

\end{center}
\end{figure*}

\cite{2014Icar..228..217T} discussed the limitations of using the quadratic equations of \cite{2010Icar..208..789D} to determine the olivine and pyroxene chemistry. They noticed that the local maxima of the equations are within the 
range of the laboratory measured Fa and Fs values of the LL chondrites. As a result, the turnover in the equations will impose an artificial upper limit in the spectrally derived iron abundances of LL chondrites. Since we have also 
used quadratic equations, the same artificial limit is expected to occur, causing in some cases an underestimation in the iron abundances of asteroids with similar compositions. However, this will not affect the classification of 
objects into this ordinary chondrite subgroup. The spectrally derived ol/(ol+px) ratios and Fa and Fs values are combined in Figures \ref{f:Figure6} and \ref{f:Figure7} for the five different wavelength ranges. In these 
figures, which have been adapted from the original work of \cite{2010Icar..208..789D}, black dashed boxes represent the range of laboratory measured values for each ordinary chondrite subgroup \citep{2010Icar..208..789D}, and 
gray solid boxes correspond to the uncertainties associated with the spectrally derived values in this work, i.e., 0.04 for the ol/(ol+px) ratio and 2.0 and 1.4, for Fa and Fs, respectively. 

\begin{table}
\begin{center}
\caption{\label{t:Table1} {\small Spectral calibrations derived from the decreased S/N spectra for the five different wavelength ranges.}}

\begin{tabular}{|c|c|c|c|c|}
\tableline

Equation No.&Wavelength Range ($\mu$m)&Spectral Calibration&R$^{2}$&rms \\ \hline

(1)&0.7-2.50&ol/(ol+px)=-0.2053xBAR+0.709&0.57&0.04 \\
(2)&0.7-2.45&ol/(ol+px)=-0.2374xBAR+0.7178&0.63&0.04 \\
(3)&0.7-2.40&ol/(ol+px)=-0.2588xBAR+0.7151&0.61&0.04 \\
(4)&0.8-2.45&ol/(ol+px)=-0.1485xBAR+0.6937&0.48&0.04 \\
(5)&0.8-2.40 &ol/(ol+px)=-0.1667xBAR+0.6959	&0.50&0.04 \\
(6)&0.7-2.40,2.45,2.50&Fa=-1283.4x(BIC$^{2}$)+2609.5x(BIC)-1295.8&0.73&2.0 \\
(7)&0.8-2.40,2.45&Fa=-1002.5x(BIC$^{2}$)+2066.5x(BIC)-1034.2&0.74&	2.0 \\
(8)&0.7-2.40,2.45,2.50&Fs=-904.4x(BIC$^{2}$)+1837.3x(BIC)-907.7&0.73&1.4 \\
(9)&0.8-2.40,2.45&Fs=-717.1x(BIC$^{2}$)+1475.3x(BIC)-733.3&0.73&1.4 \\

\tableline
\end{tabular}
\end{center}
\end{table}


\subsection{The effect of decreasing the S/N}

\begin{figure*}[!ht]
\begin{center}
\includegraphics[height=10cm]{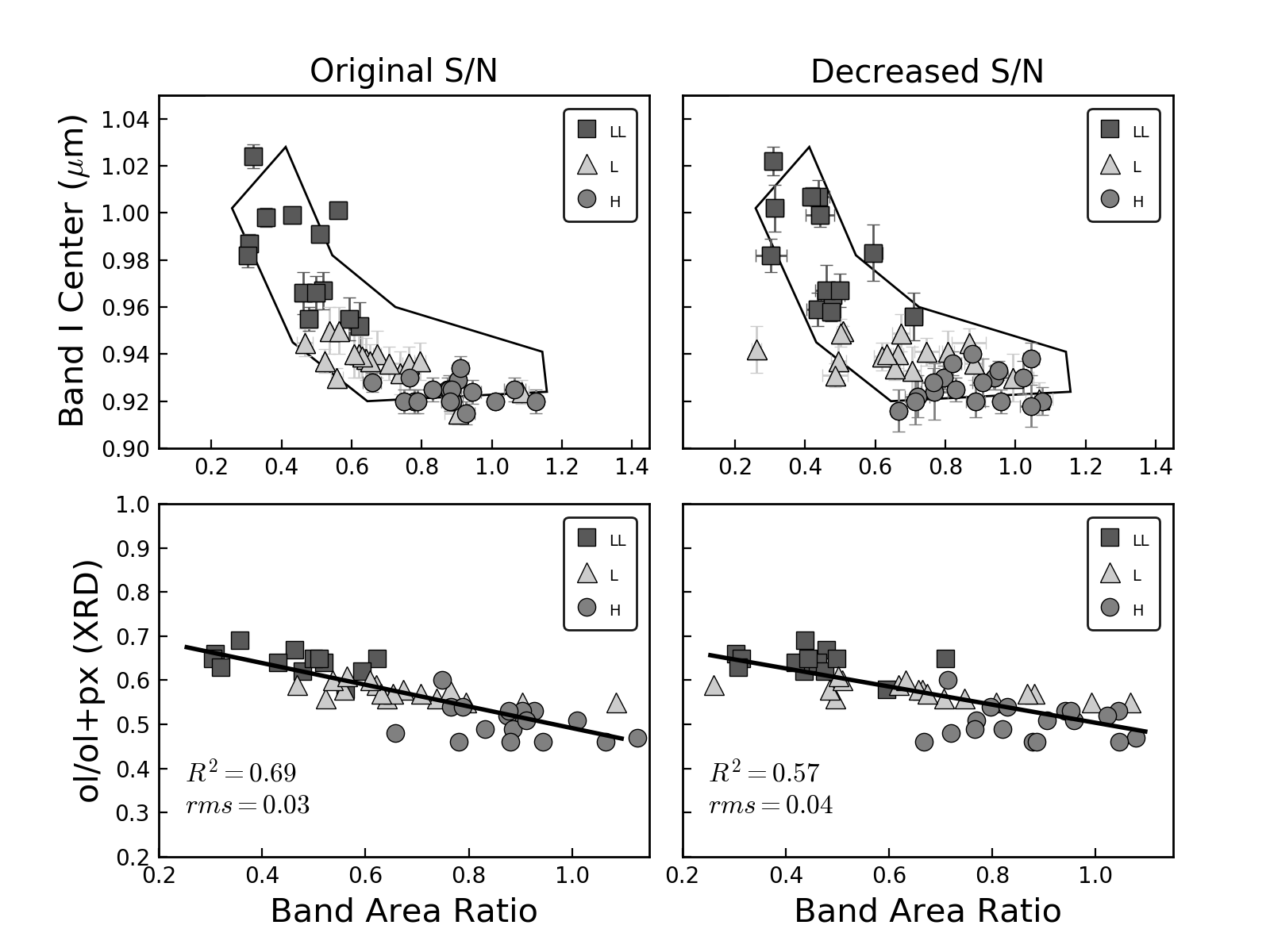}

\caption{\label{f:Figure8} {\small Top panels: Band I center vs. BAR measured from the original spectra (left) compared to those measured from the spectra with S/N $\sim$ 50 (right) for the 0.7-2.50 $\mu$m range. 
Bottom panels: XRD-measured ol/(ol+px) ratio vs. BAR corresponding to the original spectra (left) compared to those corresponding to the spectra with S/N $\sim$ 50 (right). The R$^{2}$ and rms values for each linear 
fit are also shown.}}

\end{center}
\end{figure*}

As explained earlier, the first step in our analysis was to decrease the S/N of the laboratory spectra in order to recreate the S/N observed among asteroid spectral data, and thus obtain more realistic results. We found that the 
major effect of doing this was an overall decrease in the R$^{2}$, and an increase of the rms values of the new spectral calibrations. In other words, the lower R$^{2}$ and higher rms are a consequence of a greater point-to-point 
scatter in the data resulting from spectra with a much lower S/N. This can be seen in Figure \ref{f:Figure8}, where we compare the Band I center and BAR measured from the original spectra with those measured from the spectra 
with lower S/N (top panels) for the 0.7-2.50 $\mu$m wavelength range. The XRD-measured ol/(ol+px) ratio vs. BAR are shown in the bottom panels. The R$^{2}$ corresponding to the original data is higher than the value obtained 
from the noisy spectra and closer to the R$^{2}$ (0.73) obtained by \cite{2010Icar..208..789D}.

\begin{figure*}[!ht]
\begin{center}
\includegraphics[height=10cm]{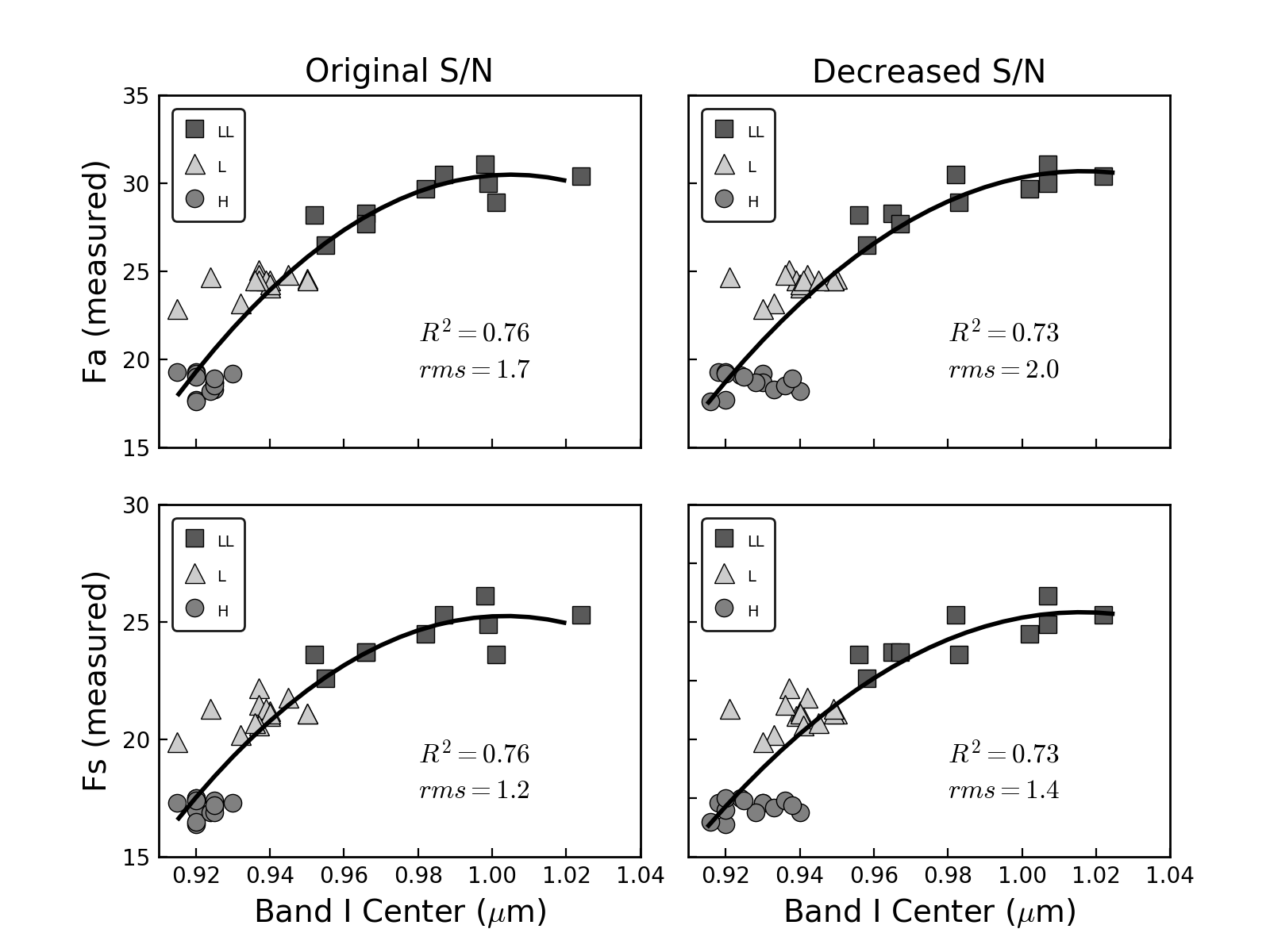}

\caption{\label{f:Figure9} {\small Top panels: measured mol\% of fayalite (Fa) vs. Band I center for the original spectra (left) and the spectra with S/N $\sim$ 50 (right) for the 0.7-2.50 $\mu$m range. Bottom panels: measured mol\% 
of ferrosilite (Fs) vs. Band I center for the original spectra (left) and the spectra with S/N $\sim$ 50 (right). The R$^{2}$ and rms values for each fit are indicated.}}

\end{center}
\end{figure*}

The decrease in S/N was also found to produce a general shift of the Band I center to longer wavelengths, being more pronounced for the LL and H chondrites. Figure \ref{f:Figure9} shows measured Fa and Fs values as function 
of the Band I center for the original spectra (left), and the noisy spectra (right) for the 0.7-2.50 $\mu$m range. This shift in Band I centers will make it harder to differentiate between L and H chondrites, which explains the tendency 
for the H chondrites to group in the upper part of their box (Figures \ref{f:Figure6} and \ref{f:Figure7}), overlapping with some L chondrites. The difference in R$^{2}$ and rms between the original and decreased S/N was found to 
be small for the five wavelength ranges.

\break

\subsection{The effect of changing the short wavelength edge}

\cite{2016M&PS...51..806L} highlighted the importance of matching how Bands I and II are defined to the methods used to derive calibration equations. Using the same 48 ordinary chondrites used in this study, and the  
\cite{2010Icar..208..789D} study, they examined how frequently a long wavelength edge, dubbed the “red edge,” of Band II set to 2.40 $\mu$m and 2.45 $\mu$m resulted in ol/(ol+px) values that were larger than the rms error from 
the meteorite calibration equations. When using a 2.40 $\mu$m red edge compared with a 2.50 $\mu$m red edge, they observed that changes in the BAR ratio resulted in ol/(ol+px) determinations outside of the rms error of the 
calibration equation for 41.67\% of the ordinary chondrite meteorites overall. For each subgroup, the choice of the red edge was the largest source of error for 72.2\% of the H chondrites, 29.4\% of the L chondrites, and 15.4\% of the 
LL chondrites. 

\begin{figure*}[!ht]
\begin{center}
\includegraphics[height=10cm]{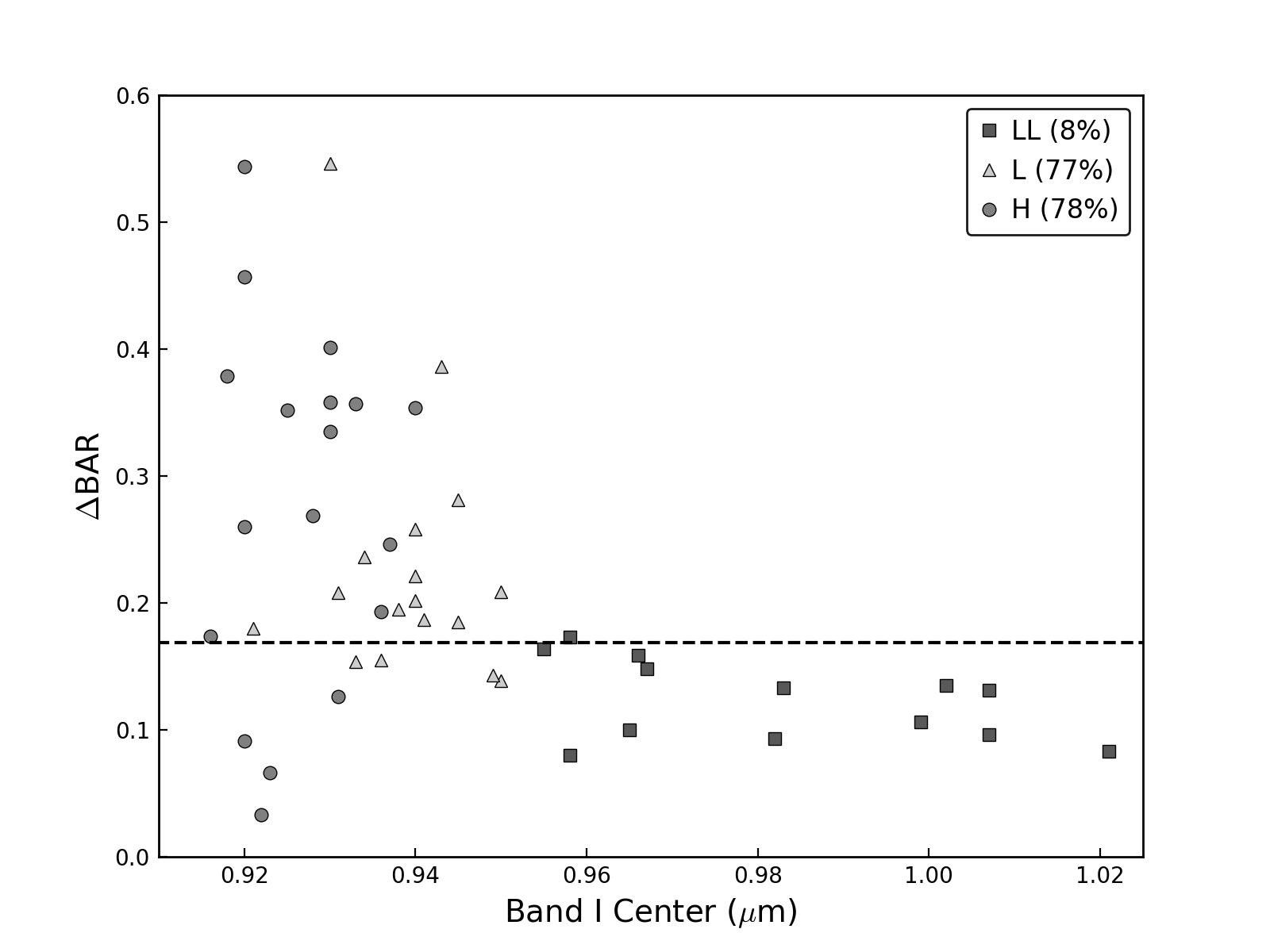}

\caption{\label{f:Figure10} {\small $\Delta$BAR as a function of the Band I center for the three subtypes of ordinary chondrites. The $\Delta$BAR is given by the difference between the BAR values measured at the two different blue 
edges (0.7-0.8 $\mu$m), while keeping the long wavelength end fixed at 2.45 $\mu$m. The dashed line corresponds to the critical $\Delta$BAR for the blue edge (see the text). Meteorites that fall above this line have $\Delta$BARs 
greater than the error inherent in Equation 2. The percentage of time that the 0.8 $\mu$m blue edge is a problem for each subtype of ordinary chondrite is indicated. Figure adapted from \cite{2016M&PS...51..806L}.}}

\end{center}
\end{figure*}

In order to investigate the extent to which changing the short wavelength edge (blue edge) is a problem, we have performed a similar analysis to the one employed by \cite{2016M&PS...51..806L} with the red edge. In our case, the 
difference lies in that we quantify the variation in BAR when the blue edge is changed from 0.7 to 0.8 $\mu$m, while keeping the long wavelength end fixed at 2.45 $\mu$m. The difference between the BAR values measured at the 
two different blue edges is given by the $\Delta$BAR, which is plotted as a function of the Band I center in Figure \ref{f:Figure10}. \cite{2016M&PS...51..806L} defined a critical $\Delta$BAR in order to identify those cases where the 
error associated with the red edge choice was higher than the intrinsic calibration error associated with the equation to estimate the ol/(ol+px) ratio derived by \cite{2010Icar..208..789D}. We have defined a similar critical 
$\Delta$BAR for the blue edge, which can be calculated from the linear equations used to estimate the ol/(ol+px) ratio from the BAR (Equations (1)-(5) in Table 1). For this particular example, where we are keeping the long 
wavelength end fixed at 2.45 $\mu$m, we use equation 2 of Table 1. The critical $\Delta$BAR is then calculated as $\Delta$BARcrit = 0.04/0.2374 = 0.1685, where 0.04 is the rms error, and 0.2374 is the coefficient in equation 2. 
The idea with this exercise is to identify those cases where the uncertainty associated with the blue edge choice (i.e., 0.7 or 0.8 $\mu$m) is higher than the intrinsic calibration error in equation 2. The $\Delta$BARcrit is depicted in 
Figure \ref{f:Figure10} as a dashed line; meteorites that fall above this line have $\Delta$BARs greater than the error inherent in equation 2. We found that for LL chondrites the percentage of time that the 0.8 $\mu$m blue edge is 
problematic is ~8\%, whereas for L and H chondrites is 77, and 78\%, respectively.  These results highlight the necessity to derive new equations to account for this problem.


\subsection{Discussion}

\cite{2014Icar..237..116R} verified the validity of the equations of \cite{2010Icar..208..789D} comparing the spectrally derived olivine and pyroxene chemistry of near-Earth asteroid (25143) Itokawa with those measured from samples 
returned by the Hayabusa spacecraft. They found a difference of less than 1 mol\% between the spectrally derived Fa and Fs values and the laboratory measurements. In this section we test if the new equations derived from 
spectra with a much lower S/N, using higher polynomial orders, and different wavelength ranges are still capable of producing similar results. For this, we measured the spectral band parameters from the NIR spectrum of Itokawa 
obtained by \cite{2001M&PS...36.1167B} with the IRTF. Band centers and the BAR were measured following the same procedure used with the ordinary chondrite spectra. A temperature correction was applied to the BAR in order to 
account for the difference between the room temperature at which the equations were derived and the lower surface temperature of the asteroid \citep[e.g.,][]{2012Icar..220...36S}. Equations (1)-(5) (Table 1) were then used to 
calculate the ol/(ol+px) ratio for the five wavelength ranges. The Band I center was used with equations (6)-(9) to calculate the mol\% of Fa and Fs. Spectral band parameters, Fs, Fa, and ol/(ol+px) values calculated for Itokawa 
are shown in Table 2. 

\begin{figure*}[!ht]
\begin{center}
\includegraphics[height=11cm]{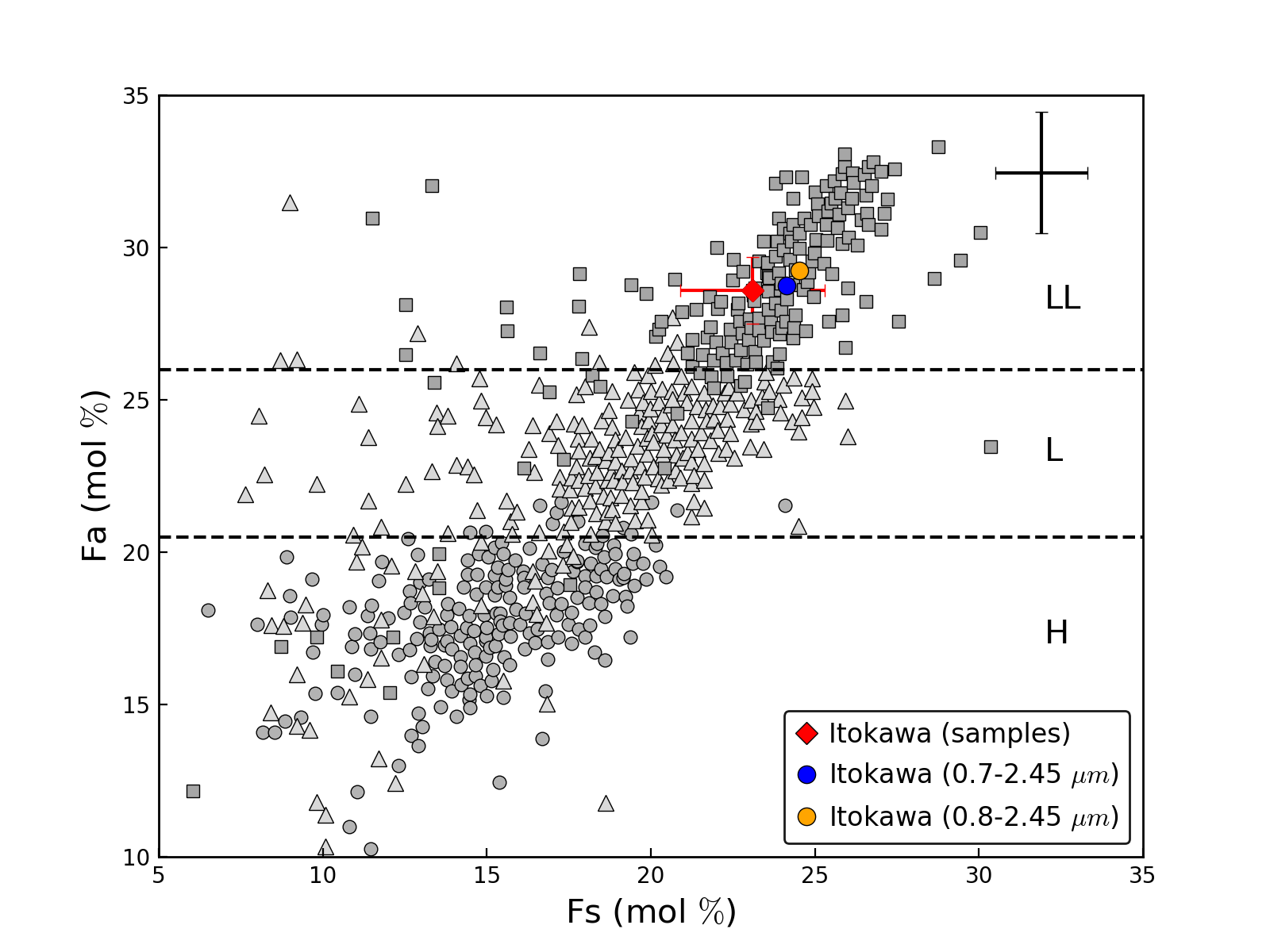}

\caption{\label{f:Figure11} {\small Spectrally derived Fa vs. Fs values for Itokawa calculated for two different wavelength ranges. Laboratory measurements of the returned samples obtained by \cite{2011ScienceNakamura} are 
shown as a red diamond. Measured values for LL (squares), L (triangles), and H (circles) ordinary chondrites are also included. The error bars in the upper right corner correspond to the uncertainties associated to the spectrally 
derived values from the model, 2.0 mol\% for Fa, and 1.4 mol\% for Fs. Figure adapted from \cite{2011ScienceNakamura}.}}

\end{center}
\end{figure*}

\begin{table}
\begin{center}
\footnotesize
\caption{\label{t:Table2} {\small Spectral band parameters and composition for asteroid Itokawa for the five different wavelength ranges. The olivine and pyroxene chemistries measured from the returned samples are 
Fa$_{28.6\pm1.1}$ and Fs$_{23.1\pm2.2}$, respectively \citep{2011ScienceNakamura}, and the ol/(ol+px) ratio is 0.76$\pm$0.10 \citep{2014M&PS...49..172T}.}}

\hspace*{-2.8cm}
\begin{tabular}{|c|c|c|c|c|c|c|}
\tableline

Wavelength Range ($\mu$m)&Band I Center ($\mu$m)&BAR&Temp. corrected BAR&Fa (mol\%)&Fs (mol\%)&	ol/(ol+px) \\ \hline

0.7-2.50&0.978$\pm$0.005&0.40$\pm$0.05&0.38$\pm$0.05&28.7$\pm$2.0&24.1$\pm$1.4&0.63$\pm$0.04 \\
0.7-2.45&0.978$\pm$0.005&0.38$\pm$0.05&0.36$\pm$0.05&28.7$\pm$2.0&24.1$\pm$1.4&0.63$\pm$0.04 \\
0.7-2.40&0.978$\pm$0.005&0.35$\pm$0.05&0.32$\pm$0.05&28.7$\pm$2.0&24.1$\pm$1.4&0.63$\pm$0.04 \\
0.8-2.45&0.992$\pm$0.005&0.55$\pm$0.05&0.53$\pm$0.05&29.2$\pm$2.0&24.5$\pm$1.4&0.62$\pm$0.04 \\
0.8-2.40&0.992$\pm$0.005&0.51$\pm$0.05&0.49$\pm$0.05&29.2$\pm$2.0&24.5$\pm$1.4&0.61$\pm$0.04 \\

\tableline
\end{tabular}
\end{center}
\end{table}

For the three wavelength ranges encompassed between 0.7 and 2.50 $\mu$m, the olivine and pyroxene chemistries of Itokawa were found to be Fa$_{28.7\pm2.0}$ and Fs$_{24.1\pm1.4}$. Higher values were obtained for 
the wavelengths in the range of 0.8-2.45 $\mu$m, Fa$_{29.2\pm2.0}$ and Fs$_{24.5\pm1.4}$, but in both cases within the uncertainties. These results are very close to the mean values measured from the samples 
(Fa$_{28.6\pm1.1}$ and Fs$_{23.1\pm2.2}$) by \cite{2011ScienceNakamura}. Spectrally derived olivine and pyroxene chemistries for Itokawa and laboratory measurements are shown in Figure \ref{f:Figure11}.

No significant variation was observed for the calculated ol/(ol+px) ratios for the five wavelength ranges; the lowest value was found to be 0.61$\pm$0.04 (0.8-2.40 $\mu$m range) and the highest was 0.63$\pm$0.04 for the three 
wavelength ranges encompassed between 0.7 and 2.50 $\mu$m. These results are similar to the one derived by \cite{2013Icar..222..273D}, who obtained an ol/(ol+px) ratio of 0.60$\pm$0.03 for Itokawa using the original equation. 
We noticed, however, that our spectrally derived values are lower than the ol/(ol+px) ratio measured from the samples by \cite{2014M&PS...49..172T} (0.76$\pm$0.10). This difference could be attributed to the limited amount of 
sample returned from Itokawa. Contrary to the disk-integrated spectrum, which represents a large fraction of the object, modal abundances measured from a few regolith particles of Itokawa might not be representatives of the entire 
asteroid. In addition, it is also possible that among the 48 ordinary chondrites used in this study, there were not enough olivine-rich LL chondrites. The highest XRD ol/(ol+px) ratio measured for the ordinary chondrites 
(0.69$\pm$0.03) corresponds to the LL6 chondrite Karatu \citep{2010M&PS...45..135D}. Thus, adding more LL chondrites with higher ol/(ol+px) ratios would increase the slope of the linear regressions in Figure \ref{f:Figure4}, 
increasing with this the spectrally derived ol/(ol+px) ratios of asteroids with a similar composition. Future studies could benefit from adding more samples, and in particular, more olivine-rich LL chondrites.  

The example presented in this section demonstrates that the equations derived for the different wavelength ranges yield similar results. The user has to decide which equations are the most appropriate for a given 
data set. For example, prior to 2017 it was a common practice to use a 0.8 $\mu$m dichroic filter with the SpeX instrument on the IRTF, as a result, hundreds of asteroid spectra were truncated at this wavelength (e.g., the 
MITHNEOS data set; \citep{2019Icar..324...41B}). In a case like this, the equations derived for the 0.8-2.45 $\mu$m wavelength range have to be applied. On the contrary, if the spectra include the local maximum at 
$\sim$0.74 $\mu$m, then the equations derived for the 0.7-2.50 $\mu$m wavelength range should be used, since they seem to yield slightly more accurate results as seen in Figure \ref{f:Figure11}. As for the long wavelength end of 
the spectrum, a visual inspection could help to determine which cutoff (2.40, 2.45, or 2.5 $\mu$m) to apply depending on the scattering of the data at these wavelengths.

\subsection{Summary}\label{sec:summ}

In this study we have complemented the original work of \cite{2010Icar..208..789D} by deriving new spectral calibrations that can be used to determine the mineral composition and abundance of ordinary chondrite-like S-type 
asteroids, i.e., objects that fall in the S(IV) compositional subgroup of \cite{1993Icar..106..573G}. Our study, which makes use of the same sample consisting of 48 ordinary chondrites, has two major differences with respect to 
the work of \cite{2010Icar..208..789D}. The first difference is that we have decreased the S/N of the laboratory spectra from $\sim$600 to $\sim$50, in order to recreate the S/N typically observed among asteroid spectra. This step 
allowed us to obtain more realistic results in terms of the uncertainties associated with the new spectral calibrations. The second difference is that the new spectral calibrations were derived for five different wavelength ranges, 
allowing us to extend the work of \cite{2010Icar..208..789D}, so the composition of the asteroids can be estimated from incomplete data.

From our analysis we found that Band I centers are shifted to shorter wavelengths compared to the values measured by \cite{2010Icar..208..789D}. The shift in Band I center arises when the polynomial order used to calculate this 
parameter is higher than the second-order used by \cite{2010Icar..208..789D}. Therefore, the new spectral calibrations are more suitable if higher polynomial orders are needed to obtain a better fit of the absorption band. If a 
second-order polynomial is enough, then the original equations of \cite{2010Icar..208..789D} should be used. Ultimately, if the procedure used to measure the band parameters is too different from the one employed in the present 
study and in \cite{2010Icar..208..789D}, new equations should be derived using that procedure.

As expected, the decrease in the S/N of the laboratory spectra caused a greater point-to-point scatter in the data, resulting in an overall decrease in the R$^{2}$, and an increase of the rms values of the new spectral calibrations. 
The decrease in S/N was also found to produce a shift of the Band I center to longer wavelengths, producing more overlap between L and H chondrites. 

We found that changing the blue edge of the spectra from 0.7 to 0.8 $\mu$m will produce variations in the BAR, which for most L and H chondrites are higher than the intrinsic calibration error. These results highlight the importance 
of deriving new spectral calibrations for different wavelength ranges.   

We tested the new spectral calibrations using the band parameters measured from the NIR spectrum of asteroid Itokawa, and comparing the results with laboratory measurements of the returned samples. We found that the 
spectrally derived olivine and pyroxene chemistry are in excellent agreement with the mean values measured from the samples. The derived mineral abundance, however, was found to be lower than the samples. This discrepancy 
could be related to the limited amount of regolith particles returned from Itokawa, which might not be representatives of the entire asteroid.

\acknowledgments

This research work was supported by NASA Solar System Workings Grant NNX16AG07G (PI: Thomas) and NASA Near-Earth Object Observations Grant NNX17AJ19G (PI: Reddy). The authors would like to thank Tasha Dunn for 
her review, which helped to improve the manuscript.

\software{NumPy (van der Walt et al. 2011), PyAstronomy (Czesla et al. 2019)}



\end{document}